\documentclass[12pt,preprint]{aastex}
\shorttitle{Be Star Disk Radii}
\shortauthors{Grundstrom \& Gies}
\begin{document}


\title{Estimating Be Star Disk Radii using \\
H$\alpha$ Emission Equivalent Widths} 

\author{Erika D. Grundstrom and Douglas R. Gies}
\affil{Center for High Angular Resolution Astronomy and \\
 Department of Physics and Astronomy,\\
 Georgia State University, P. O. Box 4106, Atlanta, GA 30302-4106; \\
 erika@chara.gsu.edu, gies@chara.gsu.edu}

\slugcomment{V3. Submitted to ApJL}
\paperid{ApJL \#20866}


\begin{abstract}

We present numerical models of the circumstellar disks of Be 
stars, and we describe the resulting synthetic H$\alpha$ emission
lines and maps of the wavelength-integrated emission flux
projected onto the sky.  We demonstrate that there are
monotonic relationships between the emission line equivalent
width and the ratio of the angular half-width at half maximum
of the projected disk major axis to the radius of the star.
These relationships depend mainly upon the temperatures of the disk
and star, the inclination of the disk normal to the line of
sight, and the adopted outer boundary for the disk radius.
We show that the predicted H$\alpha$ disk radii are consistent with
those observed directly through long baseline interferometry
of nearby Be stars (especially once allowance is made for disk 
truncation in binaries and for dilution of the observed H$\alpha$
equivalent width by continuum disk flux in the $V$-band).

\end{abstract}



\section{H$\alpha$ Spectral and Spatial Emission Models} 

Be stars are rapidly rotating B-type stars that lose mass into 
an equatorial, circumstellar disk \citep{por03}.  The disk flux 
is observed in emission lines (especially the hydrogen 
H$\alpha$ line), an infrared excess, and the 
polarization of starlight scattered in the disk.  It is now
possible to investigate the actual extent and geometry of the 
circumstellar disks through optical long baseline interferometry 
of nearby Be stars \citep{ste95,qui97,che05,tyc05,tyc06}.  
Most of these investigations have focused on the angular 
resolution of the disks in a narrow spectral band centered on the 
H$\alpha$ emission line.  The long baseline interferometric 
observations are limited to the nearby sample of Be stars 
at present, so we need other indirect methods to help 
estimate the sizes of Be star disks in more distant targets. 

Here we explore how the disk radius may be estimated 
from measurements of the disk H$\alpha$ emission flux. 
We begin with an idealized case to illustrate some 
of the important parameters that define the relationship
between equivalent width and radius.  Suppose that the 
disk is isothermal and geometrically thin    
and that the line emission is dominated by an inner
optically thick region that appears as an ellipse 
projected onto the plane of the sky.  We also make the
simplifying assumption that the specific intensity of the disk 
emission $I_\lambda^d$ is isotropic.  
Then the wavelength integrated H$\alpha$ flux will 
depend on the product of the surface flux, the wavelength
range where H$\alpha$ is optically thick, and the  
projected solid angle,     
\begin{equation}
F({\rm H}\alpha) = \pi I_\lambda^d <\triangle\lambda > \cos i \left({R_d \over D}\right)^2
\end{equation}
where $<\triangle\lambda>$ is the wavelength interval 
over which the line is optically thick (averaged over the 
visible disk), $i$ is the disk normal inclination angle, 
$R_d$ is the boundary radius of the optically thick disk, 
and $D$ is the distance to the star.   
We measure this flux through the emission 
equivalent width $W_\lambda$, which is given in 
wavelength units relative to the local continuum flux, 
\begin{equation}
F({\rm H}\alpha) = |W_\lambda| (1+\epsilon) \pi I_\lambda^s \left({R_s \over D}\right)^2   
\end{equation}
where $\epsilon$ is the ratio of disk continuum flux to 
stellar flux in the H$\alpha$ region, 
$I_\lambda^s$ is the stellar specific intensity near H$\alpha$ (assumed isotropic),  
and $R_s$ is the radius of the star.  We can then 
equate these two expressions to find the predicted relationship
between the ratio of disk to stellar radius and H$\alpha$ 
emission equivalent width, 
\begin{equation}
{R_d \over R_s} = \sqrt{ {I_\lambda^s \over I_\lambda^d}
 {{W_\lambda (1+\epsilon)}\over {<\triangle\lambda> \cos i}}}. 
\end{equation}

This special case is characterized by a disk radius that 
varies as the square root of the emission equivalent width, 
and this expectation is supported by 
the interferometric observations of \citet{tyc05} 
who find that the emitted H$\alpha$ flux scales with 
the square of the disk radius.   However, the coefficient 
of the relationship also depends on the ratio of the stellar 
to disk specific intensities that is dependent upon the 
stellar and disk temperatures.  Furthermore, there is 
an inclination dependence in both the $\cos i$ projection 
factor and the $<\triangle\lambda>$ term (through the 
range in velocities and Doppler shifts viewed through 
the disk).  The $<\triangle\lambda>$ term also depends upon the 
assumed disk density since the optical depth is based on the 
product of neutral hydrogen density and the emission line
profile.  Finally, there must also be some emission 
contribution from the optically thin outer disk that is 
ignored in expression above.  Thus, in order to predict 
accurately the relationship between equivalent width and 
disk radius, we need a more complete model that accounts 
for the specific dependences on disk temperature, inclination, 
and density. 

We have adopted the Be disk model approach of \citet{hum00}
that is based upon models developed by \citet{hor86} and 
\citet{hor95} for accretion disks in cataclysmic variables. 
The disk is assumed to be axisymmetric and centered over 
the equator of the underlying star, and the gas density varies as 
\begin{equation}
\rho (R,Z) = \rho_0 R^{-n} \exp \left[-{1\over2}\left({Z\over{H(R)}}\right)^2\right]
\end{equation}
where $R$ and $Z$ are the radial and vertical cylindrical 
coordinates (in units of stellar radii), $\rho_0$ is the base 
density at the stellar equator, $n$ is a radial density 
exponent, and $H(R)$ is the disk vertical scale height.
The neutral hydrogen population within the disk is found 
by equating the photoionization and recombination rates \citep{gie06}.  
The disk gas is assumed to be isothermal and 
related to the stellar effective temperature $T_{\rm eff}$ 
by $T_d = 0.6 T_{\rm eff}$ \citep{car06}.    

The numerical model represents the disk by a large grid 
of azimuthal and radial surface elements, and the equation 
of transfer is solved along a ray through the center of 
each element according to 
\begin{equation}
I_\lambda = S_\lambda^L (1 - e^{-\tau_\lambda}) + 
            I_\lambda^s e^{-\tau_\lambda}
\end{equation}
where $I_\lambda$ is the derived specific intensity, 
$S_\lambda^L$ is the source function for the disk gas 
(taken as the Planck function for the disk temperature $T_d$), 
$I_\lambda^s$ is the specific intensity for a uniform 
disk star (taken as the product of the  
Planck function for $T_{\rm eff}$ and an H$\alpha$ photospheric 
absorption line derived from the grid of \citealt{mar05}), 
and $\tau_\lambda$ is the integrated optical depth along
the ray (for assumed Keplerian rotation; see eq.\ [4] in 
\citealt{hor95} and eq.\ [7] in \citealt{hum00}). 
The first term applies to all the disk area elements that 
are unocculted by the star (see eq.\ [12] in \citealt{hum00})
while the second term applies to all elements that 
correspond to the projected photospheric disk of the star.  
The absorption line adopted in $I_\lambda^s$ is Doppler 
shifted according to solid body rotation for the photospheric position
in a star that is rotating at $90\%$ of the critical value. 

The model code calculates a synthetic line profile over the 
range from $-2000$ to $+2000$ km~s$^{-1}$ at 10 km~s$^{-1}$
intervals by summing the product of projected area and 
specific intensity over the disk grid.   We then integrate
the line profile relative to the continuum 
to form a predicted equivalent width (taken as
positive for net absorption or negative for net emission). 
The model also forms a synthetic image of the star plus disk 
in the plane of the sky by summing the intensity over a 
2.8~nm band centered on H$\alpha$.  We collapsed this image along the 
projected major axis to get the summed spatial intensity, 
and then we determined the radius where the summed intensity
drops to half its maximum value  
(excluding the spatial range corresponding to photospheric 
intensity).   We adopt this half-maximum summed intensity 
radius as the effective disk radius in the following 
discussion.  

We need to specify the stellar mass and radius in the 
code in order to determine the disk Keplerian velocities, 
and we adopted these as a function of stellar effective
temperature (or spectral subtype) using the eclipsing binary 
results for B-stars from \citet{har88}.  We determined 
the relationship between model equivalent width and 
disk radius by running a sequence of models defined by
the spectral subtype of the star, the disk inclination, 
the disk radial density exponent $n$, and the selected 
value for the outer boundary of the disk.   The final 
variable for any sequence is the base density $\rho_0$, 
and we computed models over a range in $\rho_0$ that
corresponds to H$\alpha$ profiles with no visible emission 
up to strong emission cases with $|W_\lambda| > 50$~\AA . 

We show the predicted relationship between the disk radius and H$\alpha$ 
equivalent width in Figure~1 for the case of a B2~Ve star 
($T_{\rm eff} = 23100$~K) assuming $n=3.0$ and an outer boundary at $100 R_s$.  
The lower solid line shows the pole-on case ($i=0^\circ$).  At higher 
inclination angles ($i=50^\circ$ and $i=80^\circ$ for the middle 
and top lines, respectively) the projected disk area is smaller 
($\propto \cos i$), and consequently the emission equivalent 
width is smaller for a given disk radius.  We do not show the $i=90^\circ$ 
case because the shear broadening approximation for the emission
profile width breaks down for an edge-on orientation \citep{hor86}.
The symbols along each sequence indicate positions of fiducial 
base density values.  Note that all three curves show the 
expected square root dependence predicted for the simple case 
(eq.\ [3] above). 

\begin{figure}
\begin{center}
{\includegraphics[angle=90,height=12cm]{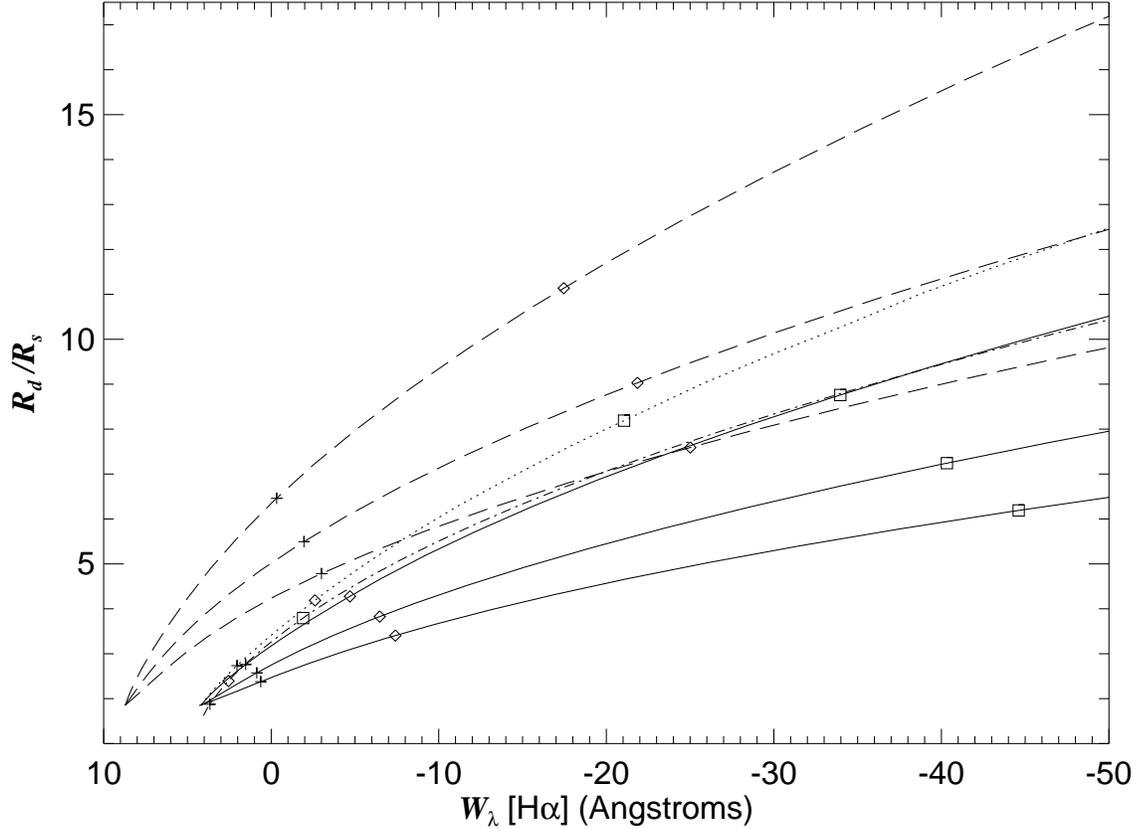}}
\end{center}
\caption{Plots of the predicted H$\alpha$ equivalent width 
and disk half-maximum radius for several model sequences.
The solid lines show the relationships for models with a B2~Ve 
central star, an outer boundary of $100 R_s$, and inclinations
of $0^\circ$ ({\it bottom}), $50^\circ$ ({\it middle}), and
$80^\circ$ ({\it top}).  The three dashed lines show the 
results from models with the same inclinations but with a 
B8~Ve star.   The dotted line shows how the relation changes 
for a model with a B2~Ve star and disk inclination of $80^\circ$
when the outer boundary is reduced to $25 R_s$.  
The dot-dashed line shows that the relation is almost identical 
for a model with a B2~Ve star, $i=80^\circ$, and $100 R_s$ boundary
when the disk density exponent is changed from $n=3.0$ to 3.5.
Disk base density values of $\rho_0=1\times10^{-12}$, $2\times10^{-12}$,
and $5\times10^{-12}$ g~cm$^{-3}$ are indicated along 
sequences by a plus sign, diamond, and square, respectively. 
}
\label{fig1}
\end{figure}

\placefigure{fig1}     

The three dashed lines in Figure~1 show the corresponding 
inclination cases for a B8~Ve star ($T_{\rm eff} = 11600$~K)  
again made assuming $n=3.0$ and an outer boundary at $100 R_s$. 
These curves are all shifted to the left relative to the 
B2~Ve sequences because the photospheric absorption component
is larger in cooler B-stars.  The model profiles of the 
B8~Ve sequence have less H$\alpha$ emission at a given 
disk radius than their B2~Ve counterparts because 
at this lower temperature the disk-to-star intensity ratio,
the vertical scale height, and the thermal emission line 
broadening are all smaller.   

The dotted line in Figure~1 shows how the H$\alpha$ equivalent 
width is reduced when the outer boundary for the disk grid 
is moved inwards from $100 R_s$ to $25 R_s$ for the case 
of a B2~Ve star with a disk inclination of $80^\circ$ 
and a disk density exponent of $n=3.0$.   
The outer disk region contains low brightness, optically thin gas, and 
although the emission contribution from any particular 
outer radius area element is small, the projected area 
of such elements increases with radius so that the 
selection of the outer boundary condition is important. 
We found that increasing the outer boundary from $100 R_s$ to 
$200 R_s$ for this parameter set increased the equivalent widths 
by only $10\%$, so we adopted the $100 R_s$ boundary as our nominal choice. 
The selection of the boundary radius will mainly
be important for Be stars in binary systems where the 
outer disk will be truncated by the gravitational influence
of the companion. 

Finally, the dot-dashed line in Figure~1 shows another 
sequence for the same B2~Ve, $i=80^\circ$, and $100 R_s$ boundary case, 
but calculated this time with a larger disk density 
exponent of $n=3.5$ (i.e., a steeper drop off in density). 
The overall shape of this curve is almost the same as 
the one for $n=3.0$ (although the fiducial base density 
points have very different locations), and this suggests 
that the particular choice of density law is not important
for the derived equivalent width -- radius relationships. 
Investigations of the infrared flux excess in Be stars 
suggest that the exponent falls in the range $n = 2 - 4$ \citep{por03}.  
We caution that the results become more sensitive to the outer disk 
boundary for $n<3$ where the emission from the optically 
thin, outer regions assumes more significance. 
 
We have constructed such model sequences for subtypes 
B0, B2, B4, B6, and B8~Ve, for inclinations $i=0^\circ$, 
$50^\circ$, and $80^\circ$, and for outer boundary 
radii of $17 R_s$, $25 R_s$, $50 R_s$, and $100 R_s$.   
These numerical results, an interpolation program, and 
our code are available to interested readers at our Web 
site\footnote{http://www.chara.gsu.edu/$^\sim$gies/Idlpro/BeDisk.tar}.


\section{Comparison with Disk Radii from Interferometry} 

We can test the predictions of the model with 
the interferometric observations of nearby Be stars 
for which the projected disk shape and size in the 
sky are known.  The long baseline interferometric 
observations in the narrow band surrounding H$\alpha$ 
are fit with an elliptical Gaussian, and the reported diameter 
$\theta_{\rm mj}^d$ is the FWHM of the Gaussian fit to the major axis.  
We list in Table~1 the observed and adopted parameters for 
the Be stars with H$\alpha$ interferometric observations
from \citet{qui97} and \citet{tyc05,tyc06}.  
The columns of Table~1 list the star name, 
a contemporaneous measurement of $W_\lambda$ and 
a code for the data source, 
adopted effective temperature \citep*{zor05}, 
the stellar angular diameter determined from 
flux fitting methods \citep{und79,och82}, 
the interferometric disk radius, ratio of the projected 
minor to major axis $r$, and a code for the data source, 
an estimate of the disk inclination, and finally 
the ratio $R_d/R_s$ from interferometry and from $W_\lambda$
(plus the adopted outer boundary for the calculation
of the latter). 
 
\placetable{tab1}      

We determined the ratio $R_d/R_s$ for each of the observations 
listed in Table~1 by interpolating in the model sequences
for $W_\lambda$, $T_{\rm eff}$, $i$, and outer disk boundary. 
We estimated the disk inclination angle by equating the 
predicted value of the projected minor to major axis 
with the observed value,  
\begin{equation}
r \approx \cos i + {C_s \over V_{\rm K}} \sqrt{R_d\over R_s} \sin i 
\end{equation}
where the disk vertical dimension is evaluated at 
radial distance $R_d$, $C_s$ is the speed of sound, 
and $V_{\rm K}$ is the Keplerian velocity at $R=R_s$
(see eq.\ [4] in \citealt{hum00}).  The ratio $C_s / V_{\rm K}$
is $\approx 0.022$ among main sequence B-stars for the 
temperatures, masses, and radii given by \citet{har88}. 
We assumed an outer disk boundary of $100 R_s$ for all 
but the known binary stars where we adopted the Roche radius 
of the primary star instead \citep{gie06,hub97}. 

The model disk radii predicted from $W_\lambda$ are 
listed in Table~1 and compared to the interferometric radii 
in Figure~2.  The predicted radii are generally 
in reasonable agreement with the observed radii, however, 
the model tends to overestimate the radii for a given 
equivalent width (or to underestimate the equivalent 
width for a given radius) for the weaker emission and cooler 
stars, $\eta$~Tau and $\beta$~CMi. 
Furthermore, the predicted radii of the stronger emission stars  
appear to be systematically lower than observed.  
We suspect that this latter problem is due to an underestimation 
of the true emission equivalent width in very active Be stars. 
If Be star disks contribute a fraction of the continuum flux in 
the $V$-band spectrum of $\epsilon=F_\lambda^d/ F_\lambda^s$, then 
we need to renormalize the equivalent widths to $(1+\epsilon) W_\lambda$
in order to compare them with the model results where the equivalent
width is referred to the stellar continuum alone.  We lack direct 
information on the value of $\epsilon$ for the targets in Table~1, 
but we can approximately estimate the renormalization factor 
from the work of \citet*{dac88} who found that 
$(1+\epsilon) \approx (1 - 0.003 W_\lambda)^{-1}$ based upon 
fits of the spectral energy distributions in a sample of Be stars
(see their eq.\ [40]).  The revised radii including this correction
({\it open squares} in Fig.~2) are in better agreement with the 
interferometric radii.  Thus, the predictions of the models appear to be 
generally consistent with the available interferometric data. 

\begin{figure}
\plotone{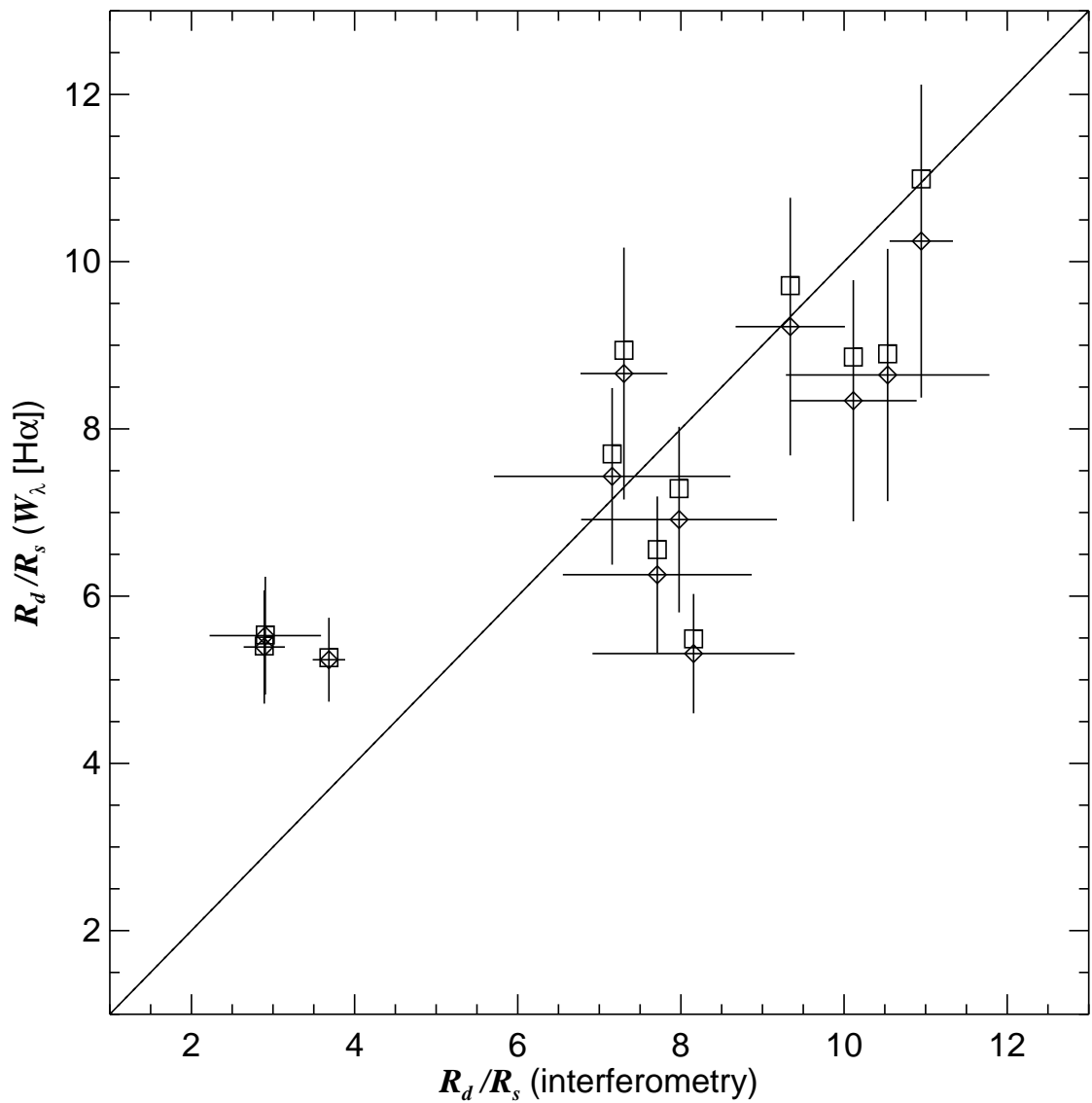} 
\caption{A comparison of the H$\alpha$ disk radii obtained 
directly from interferometry with those predicted by 
the models using the H$\alpha$ equivalent width
({\it diamonds} centered on the errors bars).  
The solid line shows the expected one-to-one correspondence.
The open squares show how the predicted
radii increase if the equivalent widths are renormalized to 
account for dilution by the disk continuum $V$-band flux
\citep{dac88}.}
\label{fig2}
\end{figure}

\placefigure{fig2}     

The method outlined above offers a new way to estimate 
disk radius for distant Be stars that is based only upon the 
H$\alpha$ equivalent width and estimates of spectral subtype, 
inclination, and outer disk boundary. 
It should prove to be a useful addition to the traditional 
method from \citet{hua72} that is based upon measurements 
of the separation between double peaks of the H$\alpha$ profile
and upon an assumed mass.  We caution readers that the derived 
radii for individual targets are probably only accurate 
to within $\pm 30\%$ at the moment (percentage standard deviation 
of the residuals for the dilution corrected radii in Fig.~2)
and that higher precision will require detailed modeling of 
the spectral energy distribution and emission lines. 
Such models will probably need a more detailed physical 
description of the disk \citep{car06} with attention paid 
to possible asymmetries in the disk gas \citep{por03}. 
The effort involved is certainly merited in studies of 
selected Be stars, but the simpler approach presented here 
may be valuable in surveys of Be stars where observations 
are limited to low resolution spectroscopic measurements of the
H$\alpha$ equivalent width (or a photometric counterpart;
\citealt{mcs05}).  The method will also be useful for 
planning future interferometric observations.


\acknowledgments

We are grateful to Dr.\ Christopher Tycner and an 
anonymous referee for their insight and comments that 
greatly aided our investigation.  
This work was supported by the National Science Foundation 
under Grant No.~AST-0205297 and AST-0506573.
Institutional support has been provided from the GSU College
of Arts and Sciences and from the Research Program Enhancement
fund of the Board of Regents of the University System of Georgia,
administered through the GSU Office of the Vice President
for Research.  




\clearpage

\begin{deluxetable}{lccccccccccc}
\rotate
\tabletypesize{\scriptsize}
\tablewidth{0pt}
\tablecaption{Be Stars with Interferometric H$\alpha$ Measurements\label{tab2}}
\tablehead{
\colhead{Star} &
\colhead{$W_\lambda$} &
\colhead{Ref.} &
\colhead{$T_{\rm eff}$} &
\colhead{$\theta^\star$} &
\colhead{$\theta_{\rm mj}^d$} &
\colhead{} &
\colhead{Ref.} &
\colhead{$i$} &
\colhead{$R_d /R_s$} &
\colhead{$R_d /R_s$} &
\colhead{$R_d /R_s$} 
\\
\colhead{Name} &
\colhead{(\AA )} &
\colhead{Col.~2} &
\colhead{(kK)} &
\colhead{(mas)} &
\colhead{(mas)} &
\colhead{$r$} &
\colhead{Col.~6, 7} &
\colhead{(deg)} &
\colhead{(interferometry)} &
\colhead{(H$\alpha$)} &
\colhead{(boundary)} 
}
\startdata
$\gamma$ Cas&   $-29.1$&1&30.2&0.45&3.47&0.70&2&49&\phn$ 7.7\pm1.2$&\phn$ 6.3\pm0.9$&\phn27\\
$\gamma$ Cas&   $-22.5$&3&30.2&0.45&3.67&0.79&3&42&\phn$ 8.2\pm1.2$&\phn$ 5.3\pm0.7$&\phn27\\
$\gamma$ Cas&   $-31.2$&4&30.2&0.45&3.59&0.58&5&58&\phn$ 8.0\pm1.2$&\phn$ 6.9\pm1.1$&\phn27\\
$\phi$ Per  &   $-35.0$&1&28.8&0.26&2.67&0.46&2&67&    $10.1\pm0.8$&\phn$ 8.3\pm1.4$&\phn22\\
$\phi$ Per  &   $-42.6$&4&28.8&0.26&2.89&0.27&5&79&    $10.9\pm0.4$&    $10.2\pm1.9$&\phn22\\
$\psi$ Per  &   $-33.7$&1&16.8&0.35&3.26&0.47&2&66&\phn$ 9.3\pm0.7$&\phn$ 9.2\pm1.5$&   100\\
$\eta$ Tau  &\phn$-6.9$&6&12.4&0.72&2.65&0.95&2&21&\phn$ 3.7\pm0.2$&\phn$ 5.2\pm0.5$&   100\\
$\eta$ Tau  &\phn$-4.3$&3&12.4&0.72&2.08&0.75&3&44&\phn$ 2.9\pm0.3$&\phn$ 5.4\pm0.7$&   100\\
48 Per      &   $-22.3$&1&16.7&0.39&2.77&0.89&2&31&\phn$ 7.2\pm1.5$&\phn$ 7.4\pm1.1$&\phn18\\
$\zeta$ Tau &   $-19.4$&6&20.1&0.43&4.53&0.28&2&78&    $10.5\pm1.2$&\phn$ 8.6\pm1.5$&\phn24\\
$\zeta$ Tau &   $-20.6$&3&20.1&0.43&3.14&0.31&3&75&\phn$ 7.3\pm0.5$&\phn$ 8.7\pm1.5$&\phn24\\
$\beta$ CMi &\phn$-3.3$&3&12.1&0.73&2.13&0.69&3&49&\phn$ 2.9\pm0.7$&\phn$ 5.5\pm0.7$&   100\\
\enddata
\tablerefs{
1. \citet{hum95};
2. \citet{qui97};
3. \citet{tyc05};
4. \citet{gie06};
5. \citet{tyc06};
6. \citet{app93}.
}
\end{deluxetable}



\clearpage

\clearpage

\end{document}